\documentclass[journal]{IEEEtran}
\usepackage{cite}
\usepackage{amsmath,amssymb,amsfonts}
\usepackage{algorithmic}
\usepackage{graphicx}
\usepackage{textcomp}
\usepackage{xcolor}
\usepackage{braket}
\usepackage{chemfig}
\usepackage{chemformula}
\usepackage{caption}
\usepackage{subcaption}
\usepackage[export]{adjustbox}
\usepackage{float}

\graphicspath{{images/}}

\usepackage{etoolbox}
\makeatletter
\patchcmd{\@makecaption}
  {\scshape}
  {}
  {}
  {}
\makeatletter
\patchcmd{\@makecaption}
  {\\}
  {.\ }
  {}
  {}
\makeatother
\def\tablename{Table}

\usepackage{hyperref}
\hypersetup{
    colorlinks=true,
    linkcolor=blue,
    filecolor=magenta,      
    urlcolor=cyan
    }

\def\BibTeX{{\rm B\kern-.05em{\sc i\kern-.025em b}\kern-.08em
    T\kern-.1667em\lower.7ex\hbox{E}\kern-.125emX}}

\title{A modular quantum-classical framework for 
simulating chemical reaction pathways accurately}

\author{
    \IEEEauthorblockN{Nirmal M R}
    , \IEEEauthorblockN{Shampa Sarkar}
    , \IEEEauthorblockN{Manoj Nambiar}
    , \IEEEauthorblockN{Sriram Goverapet Srinivasan}
    \\
    \IEEEauthorblockA{\textit{TCS Research}}\\
    \IEEEauthorblockA{\textit{Tata Consultancy Services, Mumbai, India}}\\
    \IEEEauthorblockA{mr.nirmal@tcs.com, shampa.sarkar@tcs.com}
}
\begin{document}

\maketitle

\begin{abstract}
A lot of progress has been made in recent times 
for simulating accurately the ground state energy of small 
molecules and their potential energy surface, using quantum-classical hybrid computing architecture. While these single 
point energy calculations are a significant milestone for 
quantum chemistry simulation on quantum hardware, a 
similarly important application is to trace accurately the 
reaction pathway of various chemical transformations. Such 
computations require accurate determination of the equilibrium 
or lowest energy molecular geometry, either by computing 
energy gradients with respect to molecule’s nuclear coordinates 
or perturbative distortion of the molecular configuration. In this 
work, we present a modular quantum-classical hybrid 
framework, to accurately simulate chemical reaction pathway 
of various kinds of molecular reactions. We demonstrate our 
framework by accurately tracing the isomerization pathway for 
small organic molecules. This framework can now be readily 
applied to study other ‘active’ molecules from the pharma and 
chemical industries.
\end{abstract}


\section{Introduction}
Along a chemical reaction pathway, atoms in one or more molecules rearrange themselves and their electronic or vibronic structures to transform from a set of ‘reactant’ molecules to a set of ‘product’ molecules. The pathway evolves through (at least) one unknown ‘transition state (TS)’, often characterized by a first order saddle point on the 
Born-Oppenheimer potential energy surface (BOPES) of the molecular system. Classically, predicting the transition state (TS) or simulating the reaction pathway have often been carried out by nudge elastic band (NEB) methods using ab initio techniques such as the Density Functional Theory 
(DFT) \cite{jonsson1998nudged}. For molecules with strong electronic correlations, 
DFT simulations produce inaccurate results due to 
approximate consideration of electronic exchange-correlation. NEB method also suffers from slow convergence of path optimization for complex chemical reaction, where 
dimer method has been considered \cite{kastner2008superlinearly}. Other classical 
methods comprise IRC (intrinsic reaction coordinates), which calculates the mass weighted cartesian coordinates of 
molecular systems from the transition state to the reactant and product state along the chemical reaction pathway \cite{fukui1981path}. 
However, all these quantum chemistry simulations on
classical computers are restricted to small molecules, as the cost of computing grows exponentially with system size (\(O(N^3)\) for DFT calculations, $N$ being the number of electrons).

Quantum Computing harnesses the power of quantum 
mechanical systems to perform quantum mechanical 
calculations. The technology is thus well-suited to compute the properties of physical, chemical and biological systems and the reactions therein, based on probing the electronic or vibronic degrees of freedom which are inherently quantum 
mechanical in nature. In fact, the promise held by quantum computers lies in representing the state of a system under study with fewer resources than their classical counterparts owing to the quantum mechanical effects of superposition 
and entanglement. Here we present an inhouse quantum-classical hybrid framework, called the IRC driven Variational 
Quantum Eigensolver (IRC-VQE), that simulates very 
accurately reaction pathway for small drug-like molecules, 
cis-trans (stereo) isomerization pathway for simple alkene, 
conformational isomerization pathway for simple alkane, and 
intramolecular transformations for fluxional molecules, as a 
function of their respective (varied) IRC.

Section \ref{methods} highlights the methods and the underlying theory 
briefly while the results from the application of our framework to two different molecules are discussed in section \ref{results}. 

\section{Methods- An Architectural Overview} \label{methods}

\begin{figure*}
    \centering
    \includegraphics[width=\textwidth]{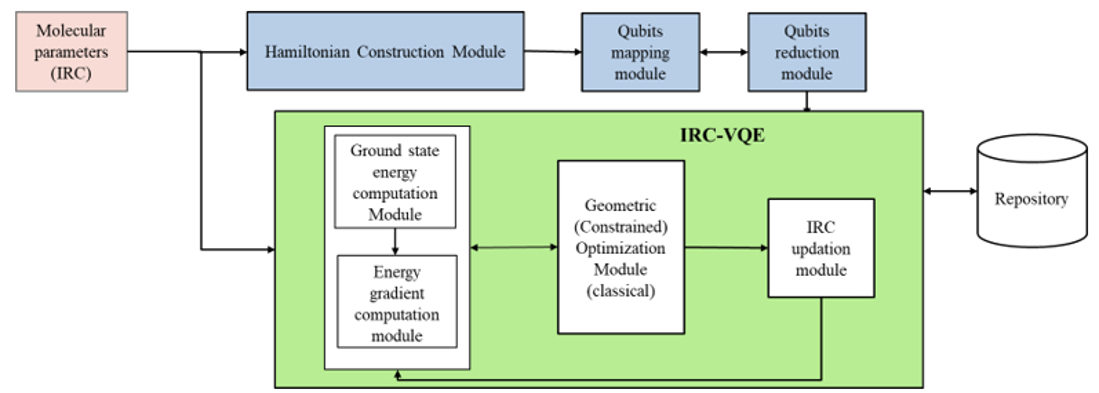}
    \caption{Architecture of IRC-VQE Framework}
    \label{fig:IRC-VQE_Framework}
\end{figure*}

The various modules constituting the quantum-classical 
hybrid framework of IRC-VQE are shown in Fig. \ref{fig:IRC-VQE_Framework}. and a 
description of how they can be efficiently realized on a Noisy 
Intermediate Scale Quantum (NISQ) device is given below:

\subsection{Molecular Parameters Module} \label{molecular parameters}

The molecular parameters module includes the  
geometry data, overall charge and spin 
multiplicity, molecular intrinsic reaction coordinates, 
point group symmetry and other relevant 
parameters associated with the molecule of interest.

The geometry of a molecule essentially refers to the arrangement of its constituent atoms in space. The most universal and widely followed representation is the Cartesian coordinate system wherein the geometry of a molecule is specified using the x, y, and z coordinates of its constituent 
atoms. However, such a representation scheme conveys little about the chemical bonding in the molecule and scales as $O(3N)$
for a molecule with N atoms. As such, this choice of representation is not effective for geometry optimization techniques as the coordinates are highly coupled to each other leading to ill-conditioned Hessian matrices, with slower convergence of the gradient-based algorithms \cite{schlegel2011geometry}.

A better choice for representing molecular structure in the context of geometry optimization and reaction path following would be the internal coordinates consisting of the bond lengths, bond angles and torsional angles. It has been observed through numerical simulations that the coupling 
between bond stretching, angle bending and torsion about single bonds is significantly smaller compared to that of Cartesian coordinates. Moreover, the ability of internal coordinates to inherently capture the bonding of the molecule 
and represent the curvilinear motions on the potential energy surfaces make them a natural choice for reaction path tracing algorithms. We can further incorporate molecular point group symmetries to reduce significantly the number of internal 
coordinates considered for simulation, that can significantly enhance the computational efficiency.

\subsection{Molecular Hamiltonian Construction Module} \label{molecular hamiltonian}

For a given geometry of a molecule, one can derive the 
Hamiltonian describing the behavior of the electrons in the 
second quantization formalism as

\begin{equation}
    \hat{\mathcal{H}} = \sum_{p,q} h_{pq}a_{p}^{\dagger}a_{q} + \frac{1}{2}\sum_{p,q,r,s} g_{pqrs}a_{p}^{\dagger}a_{q}^{\dagger}a_{r}a_{s} + E_{NN}
\end{equation}

\begin{equation}
    h_{pq} = \int dx\phi_p^*(x)(-\frac{ \nabla^{2}}{2}-\sum_I\frac{Z_I}{| {r-R_I}|})\phi_q(x)
\end{equation}

\begin{equation}
    g_{pqrs} = \int dx_1dx_2\phi_p^*(x_1)\phi_q^*(x_2)\frac{1}{|r_1-r_2|})\phi_r(x_2)\phi_s(x_1)
\end{equation}

where $h_{pq}$ and $g_{pqrs}$ are the one-electron and two-electron integrals, respectively and $E_{NN}$ denotes the nuclear repulsion energy \cite{mcardle2020quantum}. The one-body integral describes the kinetic energy of electrons and their interaction with the nuclei, whereas the two-body integral describes the electron-electron interaction. The operators $ a_{p}^{\dagger}$ and $a_{q}$ represent creation and annihilation of electron in spin orbitals $\phi_p(x)$ and $\phi_q(x)$, respectively, and obey the fermionic anti-commutation relations, where $x_i$ is the spatial and spin coordinate of the $i^{th}$ electron. 

The construction of the second quantized molecular Hamiltonian requires the computation of the two integrals in an appropriate Molecular Orbital (MO) basis set that can describe the system with a high degree of accuracy. These MOs are usually expressed as linear combinations of atom-centered basis functions and describe the spatial distribution of electrons in the molecule as well as their spin configuration. The STO-kG minimal basis sets, which approximates each Slater Type Orbital (STO) with k Gaussian Type Orbitals (GTOs) in the least squares sense, are adequate to obtain a qualitative description of the PES, equilibrium geometries, and chemical reaction paths with limited accuracy. For better quantitative understanding of the properties of chemical systems, one needs to employ larger basis sets, such as split-valence basis like 3-21G, 6-31G*, or correlation-consistent polarized basis sets like cc-pVDZ in their electronic structure calculations. A comparison of the reaction pathway for different MO basis sets have been presented in Section \ref{case study 1}.

As will be described in detail in the following Qubit Mapping module (Section \ref{qubits mapping}), for the time being, one can imagine that one spin 
orbital (SO) (spin component of a molecular orbital, since two electrons with opposite spin configuration can occupy one MO) is being mapped to one qubit of the quantum device on which the wavefunction of the molecule of interest is prepared. Hence, as one might expect, the lack of availability 
of large number of qubits in today’s NISQ devices and the requirement of shallow quantum circuits has forced a trade-off between the computational accuracy and quantum resources in the sense that choosing a larger basis set leads to 
a quantum circuit with larger width and depth.

To circumvent this trade-off, several approximate schemeshave been proposed to reduce the number of SOs in the molecular Hamiltonian. One popular strategy is to ‘freeze’ the core orbitals from the electronic structure calculation assuming that the electrons residing in the core orbitals do not 
participate in chemical bonding and consider a mean-field potential generated by the core electrons. Another well-known technique is to remove the outermost unoccupied orbitals from the calculation with the assumption that 
electrons won’t undergo transition to these high energy orbitals. Since these schemes bring in such approximations, the computational accuracy is significantly compromised, and their scalability is limited to molecules with few atoms 
and minimal basis sets. Other approaches consider 
incremental consideration of Full Configuration Interaction (iFCI), by decomposing the problem into n-body interaction terms and performing the electronic structure calculation in 
parallel for low-n systems \cite{zimmerman2019evaluation}.

A more recent approach relies on the classical-quantum embedding schemes that are based on the notion of Active Space (AS) \cite{rossmannek2021quantum}. In particular, the entire system is split into a set of active orbitals, where the electrons can interact among 
themselves, and its environment, which is comprised of the inactive orbitals with negligible interaction between the electrons \cite{sun2016quantum}. 
Orbitals near the Fermi-level or highest occupied MO/lowest unoccupied MO (HOMO/LUMO) are usually taken as active orbitals. In this embedding approach, only the active orbitals 
are mapped to qubits and are given a quantum computing treatment whereas the rest is treated with an efficient classical algorithm like Density Functional Theory (DFT) or Hartree-Fock (HF). The effective Hamiltonian incorporates a 
mean field potential generated by the inactive electrons, 
hence replaces the mapping of inactive orbitals to qubits as 
given in Equation \ref{eqn:active space hamiltonian}

\begin{equation}
    \hat{\mathcal{H}_{qc}} = \sum_{u,v} F_{uv}^{I}a_u^{\dagger}a_v + \sum_{u,v,x,y}g_{uvxy}a_u^{\dagger}a_v^{\dagger}a_xa_y
    \label{eqn:active space hamiltonian}
\end{equation}

where $F_{uv}^{I}$ is the inactive Fock operator that embeds the quantum computation into the classically computed environment. The effective Hamiltonian $\hat{\mathcal{H}_{qc}}$ takes into account only the active orbitals indexed by $u,v,x$ and $y$.

Compared to the other techniques, the advantage of this embedding strategy lies in outsourcing the calculation of inactive electrons to a classical routine, thereby replacing the mapping of inactive orbitals to qubits, making the entire 
computation highly efficient and maintaining a good degree of accuracy at the same time. This strategy also facilitates one 
to employ higher-order basis sets in the electronic structure calculation to further enhance the accuracy with optimal quantum resources.

\subsection{Qubits Mapping Module} \label{qubits mapping}

To leverage the potential of quantum computers to efficiently simulate chemical systems, one needs to transform the electronic Hamiltonian in second quantized representation to the operators that act on qubits. The most prominent encoding 
techniques to perform this mapping of a fermionic state to a qubit state and the corresponding operators are Jordan-Wigner, Parity, and Bravyi-Kitaev encodings.

One of the simplest and earliest encodings introduced in 1928 is the Jordan-Wigner (JW) mapping \cite{wigner1928paulische}. In this encoding, the 
state $(q_i)$ of a qubit $i$  encodes the occupation number $f_i$ of the corresponding spin orbital. Mathematically,
\begin{equation}
    \ket{f_{M-1}, f_{M-2},\dots,f_0} \xrightarrow[]{} \ket{q_{M-1}, q_{M-2},\dots,q_0},
\end{equation}
\begin{equation}
        q_p = f_p \in {0,1}
\end{equation}

The mappings of the fermionic creation and annihilation operators under Jordan-Wigner encoding are
\begin{equation}
    a_p = Q_p \otimes Z_{p-1} \dots \otimes Z_0
\end{equation}
\begin{equation}
    a_p^\dagger = Q_p^\dagger \otimes Z_{p-1} \dots \otimes Z_0
\end{equation}

where $Q = \frac{1}{2}(X+iY)$ and  $Q^{\dagger} = \frac{1}{2}(X-iY)$ and $X$, $Y$ and $Z$ are the single-qubit Pauli operators. Being a trivial 
mapping, JW encoding requires as many qubits as there are number of spin orbitals in the electronic Hamiltonian.

The second encoding scheme is the Parity mapping. Here, the parity information of the first $p$ spin orbitals is being encoded into the state of $p^{th}$ qubit. The mapping of creation and 
annihilation operators under this scheme is detailed in \cite{seeley2012bravyi}. The qubit operators resulting from Bravyi-Kitaev (BW) mapping are found to be more complex than those of the JW or parity encodings. Interested readers are directed to the works of \cite{kitaev2002fermionic} for an elaborated explanation.   
 
\subsection{Qubits Reduction Module} \label{qubits reduction}
 The qubit Hamiltonian obtained after JW, Parity or BK encoding methods will have the same number of qubits as there are spin orbitals in the electronic structure calculation. In the NISQ era of quantum computation, the quantum hardware imposes numerous restrictions to perform quantum simulations in the form of limited qubit count and connectivity, presence of noise in quantum gates (low gate-fidelity), shorter coherence times and many others. All these factors set an upper bound to the size of circuits that can be executed reliably on these devices \cite{preskill2018quantum}. Consequently, techniques that can enable the reduction of resources required for quantum simulation often gets widely acknowledged in the quantum community.

One such popular technique that facilitates the removal of two qubits when used with parity and BK encoding schemes, as detailed down in \cite{bravyi2017tapering}, is the method of Z2 Symmetries. This technique utilizes the fact the total number of electrons and total spin value are conserved by Hamiltonian to identify certain qubits which are acted on by the identity or Pauli Z operators only in every term of the Hamiltonian. We can then replace these operators with their eigenvalues, thus reducing the requirement of qubits by two for the simulation on a quantum device. Further improvement on this technique with the inclusion of molecular point group symmetries have been worked on by \cite{setia2020reducing}.

\subsection{Ground State Energy Computation Module (herein Variational Quantum Eigensolver} \label{VQE}

The Variational Quantum Eigensolver (VQE) is a hybrid quantum-classical algorithm that is used extensively to find the ground state energy (lowest energy) of a molecular system for a given geometry. VQE was first proposed in 2014 by Alberto Peruzzo and coworkers and experimentally implemented on 
superconducting qubits to simulate the ground state energy of small molecules in 2017 \cite{peruzzo2014variational} \cite{kandala2017hardware}. Computing ground state energy is often the first and foremost task in most quantum chemistry calculations. Knowledge of this quantity with a 
high degree of accuracy can potentially lead us to
understanding chemical reactions better and calculation of 
other molecular properties, such as time evolution of chemical systems \cite{sokolov2021microcanonical}.

\begin{figure}[!htb]
    \centering
    \includegraphics{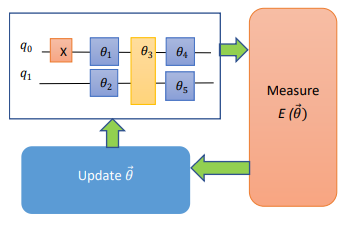}
    \caption{A simple schematic of Variational Quantum Eigensolver}
    \label{fig:VQE_schematic}
\end{figure}

VQE orchestrates a hybrid quantum-classical computing architecture, by using a quantum processing unit (QPU) for preparing the parameterized trial quantum state of the 
molecule and measuring the energy operated upon the quantum state, and a classical processing unit (CPU) to adjust the parameters of the quantum circuit or state depending on 
the measurement results from QPU. The advantage of VQE over classical methods comes from the fact that a QPU can represent and store the wavefunction of the system studied 
more efficiently than a CPU. The use of a classical optimizer to update the parameters of the quantum circuit makes VQE a better candidate than fully quantum algorithm like Quantum 
Phase Estimation (QPE) on near-term quantum hardware as it overrides the long coherence times and deeper circuits needed by QPE with measurement repetitions and shallower circuits \cite{abrams1999quantum}.

VQE is derived from the variational principle of Quantum Mechanics. According to this principle,

\begin{equation}
    \bra{\psi(\vec{\theta})}H\ket{\psi(\vec{\theta})} \geq E_0
\end{equation}
where $\ket{\psi(\vec{\theta})}$ is a parametrized trial electronic wavefunction, $E_0$
is the lowest eigenvalue of the Hamiltonian $H$ and $\vec{\theta} = (\theta_1, \theta_2, \dots, \theta_n)^T$. This immediately implies that minimizing the energy (given by the expectation value of $H$) with respect to parameters $\vec{\theta}$ will enable one to determine the ground state wavefunction $\psi(\vec{\theta})$ and energy. Since QPUs are more 
efficient in preparing and storing the trial wavefunction, one can employ a QPU for this task and the parameters can be adjusted through a classical optimization method as shown in figure \ref{fig:VQE_schematic}.

The quantum state or ‘ansatz’ is prepared on a quantum processor by means of a parametrized quantum circuit. The choice of an 
appropriate ansatz has been a matter of extensive research in the quantum community as the accuracy and efficiency of the estimated ground state energy computation primarily depends on the ansatz chosen. At a broader level, ansatzes are 
classified in two types: hardware-efficient and chemically inspired. The former class of circuits typically comprises of gates that are native to the available quantum hardware on which one intend to run VQE \cite{kandala2017hardware}. These circuits generate the trial state with as few gates as possible owing to the 
shorter coherence times and restricted gate topologies of current devices. However, one drawback of hardware-efficient ansatz is that they do not consider the details of the 
molecule being simulated and thus the solution space may not include the desired optimal solution.

The second class of ansatz derives their structure from classical computational chemistry algorithms, hence the name chemically inspired. In particular, one of the first ansatz explored is the unitary variant of the coupled cluster (CC) 
method \cite{bartlett1989alternative}. The Unitary Coupled Cluster (UCC) ansatz attempts to create the trial state by including excitations, 
typically, the single and double excitations above the reference state (UCCSD). Because of the non-truncation of the BCH series, no efficient classical implementation of the 
UCC method has been reported till date, but can be efficiently constructed on quantum processors \cite{taube2006new}. The UCC method 
inherits all the benefits of the CC method in addition to being fully variational. 

Once the trial electronic state has been constructed, we need to measure the energy associated with the state. The Quantum 
Expectation Estimation (QEE) algorithm can be used to serve this purpose on a quantum computer by repeated measurements \cite{peruzzo2014variational}. The Hamiltonian acting on the qubits can be expanded as

\begin{equation}
    \hat{\mathcal{H}} = \sum_j h_{j}P_{j} = \sum_j h_{j}\prod_i\sigma_i^j
\end{equation}

where $h_j$ is a real number, $\sigma_i^j \in {I,X,Y,Z}, i$ refers to the 
qubit being acted on, and $j$ represents an individual Pauli 
string in $\hat{\mathcal{H}}$. By exploiting the linearity of quantum observables, 

\begin{equation}
    E(\vec(\theta_k) = \sum_{j}h_j\bra{\psi(\vec{\theta_k})}\prod_{i}\sigma_{i}^j\ket{\psi(\vec{\theta_k})}
\end{equation}

Repeating these state generation and measurement steps many times for every term in the Hamiltonian gives the energy of the trial state. Then, the current parameter vector $\vec{\theta_k}$ and the energy $E(\vec{\theta_k})$ are fed to a classical optimization algorithm, which then generates a new parameter vector $\vec{\theta_{k+1}}$. The quantum circuit prepares a new trial state $\ket{\psi(\vec{\theta_{k+1})})}$ with the updated parameter vector, which has a 
lower energy than $\ket{\psi(\vec{\theta_{k}})}$. These steps are iterated till the energy reaches a minimum.

\subsection{Energy Gradient Computation Module} \label{energy gradient}
Most of the work in the recent past on using a quantum computer to address quantum chemistry problems has focused predominantly on ground state and/or excited state 
energy calculations \cite{peruzzo2014variational}\cite{kandala2017hardware} \cite{ollitrault2020quantum}. For many applications, such as geometry optimization, transition state searches, following reaction paths, molecular dynamics (MD) 
simulations, one needs to have the knowledge of derivatives of electronic energies with respect to some physical parameter. For instance, the energy derivative w.r.t nuclear 
positions gives the force acting on the nuclei, energy gradient w.r.t external electric field or magnetic field yields the electric and magnetic dipole moments, respectively.

In the literature, there exists two broad classes of methods to compute energy gradients \cite{o2021efficient}. The first class of methods relies on the famous Hellmann-Feynman theorem to derive analytical expressions for energy gradients \cite{feynman1939forces}. According to this theorem, the energy derivative with respect to a parameter is equal to the expectation value of the derivative 
of the Hamiltonian with respect to the same parameter 
\begin{equation}
    \frac{dE}{dx}=\bra{\psi}\frac{dH}{dx}\ket{\psi}
\end{equation}
where $\ket{\psi}$ is a normalized eigenstate of the Hamiltonian $H$ and $x$ is any general parameter mentioned above. In practice, one 
needs to incorporate all the extrinsic and intrinsic dependencies of $H$ in order to calculate the exact total derivative of $H$ and that can be complex. One of the proposed 
approaches to compute energy gradients in the NISQ era is outlined as follows:
\begin{enumerate}
    \item Define a set of basis rotations $\{Q_j\}$ for which shallow quantum circuits can be constructed.
    \item For each rotation $Q_j$, prepare the quantum state $\ket{psi}$ $M_j$ times, apply the quantum circuit, and destructively measure the system in the computational basis.
    \item Estimate the expectation value $\{\bra{\psi}\frac{dH}{dx}\ket{\psi}\}$
    from the set of measurements.
\end{enumerate}
Here, the aim is to reduce the total number $S=\sum_j{S_j}$ state 
preparations or measurements.

The second set of methods to compute energy gradients finds application in the era of fault-tolerant quantum computers \cite{o2021efficient}. For the sake of simplicity, we briefly describe one of the 
algorithms based on a numerical estimation of the energy gradient with respect to nuclear positions. For numerical differentiation of energy, one needs to use finite difference methods that calculate energy gradients as linear combinations of energies for different nuclear configurations 
of the molecule. Among the different finite difference techniques, central difference formula offers a quadratic advantage with respect to discretization error as compared to 
both forward and backward finite differences. The simplest central-difference expression that calculate the partial derivative of energy with respect to one of the nuclear coordinates is 
\begin{equation}
    \frac{dE}{dR_i} = \frac{E(\vec{R}+ \frac{ dR\cdot{\vec{v_i}}}{2}) -E(\vec{R}-\frac{ dR\cdot{\vec{v_i}}}{2})}{dR} + \epsilon_{fd}
\end{equation}
where $\vec{v_i}$ is the unit vector along the component $R_i$ and $\epsilon_{fd}$ is the error due to finite difference approximation. In this 
approach, the quantum computer is used as a subroutine to estimate energies at two different nuclear positions $\vec{R}+ \frac{ dR\cdot{\vec{v_i}}}{2}$ and $\vec{R}- \frac{ dR\cdot{\vec{v_i}}}{2}$. The above formula can be generalized to higher-order finite differences. These finite difference-based 
methods require further optimization on the step-size $dR$ and 
the degree of finite difference method used to lower the error $\epsilon_{fd}$.

\subsection{Geometry Optimization Module} \label{geometry optimization}
In computational chemistry, geometry optimization is the procedure that aims to determine the most stable arrangement of atoms in a given molecule. As the name suggests, this is essentially an optimization problem where the total ground state energy of a molecule is minimized with respect to the 
nuclear positions such that the net force acting on each nucleus is zero. The equilibrium geometry obtained from this calculation is often the starting point for computation of other molecular properties and for tracing chemical reaction paths 
on the PES.

Most efficient classical algorithms for molecular geometry optimization require the expensive calculation of the second order derivative of energy with respect to nuclear coordinates (the Hessian matrix) at each iteration. Since the exact 
computation of molecular ground state energy is intractable on a classical computer for molecules with more than ten electrons, the use of finite difference techniques to predict the 
gradients and Hessian become even more computationally challenging, thus invoking the use of force-fields to estimate the gradients.

\begin{figure*}
    \centering
    \includegraphics[width=\textwidth]{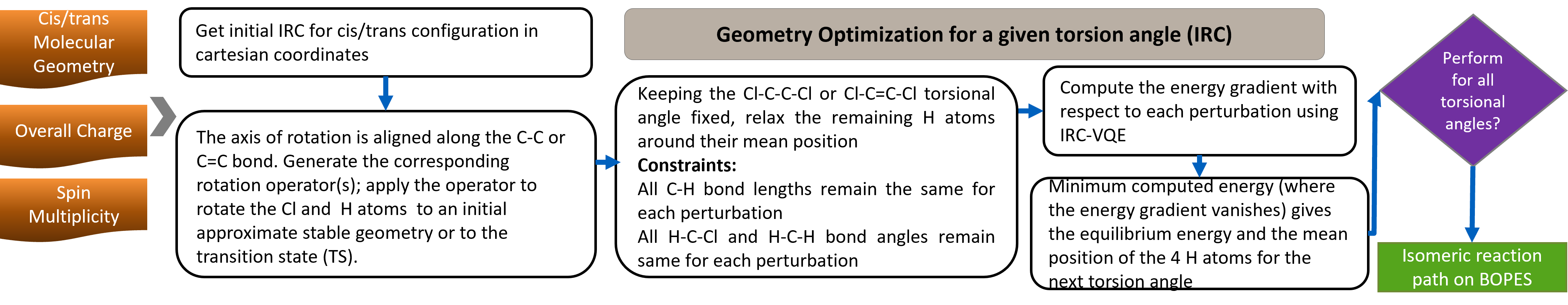}
    \caption{Block diagram of IRC-VQE applied to trace isomerization reaction pathway}
    \label{fig:Isomerization_IRC-VQE}
\end{figure*}

Under the Born-Oppenheimer (BO) approximation of 
stationary nuclei, one can parametrize the molecular Hamiltonian in terms of the nuclear coordinates and thus express the molecular energy $E(x)$ as a function of the nuclear coordinates $x$, defining the BOPES. Solving the equation $\nabla E(x)=0$ for $x$ yields the equilibrium geometry of 
the molecule and corresponds to the global minimum on the BOPES. In the NISQ era, one can opt for different methods like the perturbative method in which the atoms of the molecule are relaxed around some initial point in a constrained manner maintaining the symmetry of the molecule []. 
The advantage of this method is that it can be computationally efficient as it doesn’t require the explicit evaluation of energy gradients or Hessian and minimizes the number of variables optimized by taking into consideration the inherent symmetry of the molecule. The main drawback of this 
approach is that it becomes infeasible for a complex molecule that lacks symmetry and too many degrees of freedom come into picture. This method cannot be applied in scenarios where highly accurate calculation of optimized geometry and energy is necessary. A recent wok on geometry optimization 
by Xanadu explores a joint optimization scheme wherein they tried to optimize both the circuit parameters and nuclear coordinates at each optimization step simultaneously to reach 
the equilibrium geometry \cite{delgado2021variational}. This approach requires defining a new objective function as the expectation value of the parametrized Hamiltonian for the parametrized trial state and depends on both circuit parameters and nuclear coordinates.

\subsection{Intrinsic Reaction Coordinates (IRC) Module} \label{IRC}
In classical computational chemistry, one of the popular techniques to trace the reaction dynamics of chemical systems is known as the Intrinsic Reaction Coordinate (IRC), proposed by Fukui \cite{fukui1981path}. The IRC is defined as the minimum energy reaction pathway (MERP) on the BOPES in mass-weighted cartesian coordinates between the transition state of 
a reaction and its reactants and products. Following the IRC involves identification of the reaction coordinate of the reacting species, finding the minimum energy and corresponding optimal geometry for each reaction coordinate, and updating the reaction coordinate and initial 
geometry based on the outputs of previous reaction 
coordinate.

The IRC can be obtained by solving the following differential equation
\begin{equation}
    \frac{d\vec{q}}{ds} = \vec{v}(s)
\end{equation}
where $\vec{q}$ is the mass-weighted Cartesian coordinates, s is the coordinate along the IRC, and $\vec{v}$ is the normalized tangent vector at each point on IRC.

\begin{figure}[!htb]
    \centering
    \includegraphics[width=0.45\textwidth]{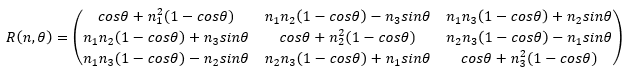}
    \caption{Matrix representation of the rotation operator}
    \label{fig:Rotation_Operator}
\end{figure}

Some of the merits of IRC include that it provides a unique connection from a given TS to the reactant and product states. Many a time, the IRC approach has been used to verify whether a given TS is connected to the reactant and product 
for a specific chemical reaction. Since bond stretching and bending requires more energy than torsional motions, in many reactions the reaction coordinate is the torsional angle formed by 4 non-colinear atoms. To facilitate the rotation of 
atoms to vary the torsional angle, one can apply a rotation operator that can be represented in matrix form as given in Fig. \ref{fig:Rotation_Operator}
where the axis of rotation is defined by a unit vector with components $n1, n2, n3$ and $\theta$ is the angle of rotation.

The flowchart depicting the entire approach to simulate isomerization pathways for alkanes and alkenes using IRC-VQE is shown in Fig.  \ref{fig:Isomerization_IRC-VQE}.

\section{Results and Discussion} \label{results}
We present the results obtained with the numerical simulation of two molecules, namely 1,2-dichloroethylene (\ch{C2H2Cl2}) and ethane (\ch{C2H6}), using the IRC-VQE algorithm.

\subsection{Case Study 1: IRC-VQE applied to \ch{C2H2Cl2}} \label{case study 1}

\ch{C2H2Cl2} is an example of the simplest organic molecules that contain a carbon-carbon double bond. The molecule exists in either of the two geometric isomeric configurations, cis-1,2-dichloroethene,  
which has two Chlorine (Cl) atoms on the same side of the double bond, and, trans-1,2-dichloroethene, which has two Cl atoms on the opposite side of the double bond.

Molecules like \ch{C2H2Cl2} with a double bond between the C atoms are good candidates for drug-like molecules and knowing and being able to compute the isomerization energy landscape of similar alkene compounds is of substantial 
relevance to the pharmaceutical industry for drug discovery. Towards this end, in 2020, the Google Quantum AI team has simulated two isomeric reaction pathways for diazene, a similar molecule, but with a nitrogen-nitrogen double bond \cite{google2020hartree}. They claimed that their Sycamore quantum processor was able to resolve the energy barrier of 40 milli-Hartree 
between the two transition states.

\ch{C2H2Cl2} molecule has a planar structure in both isomeric forms, meaning all the atoms are in a single plane. The conversion from cis- form to the trans-form requires the exchange of positions of Cl and H atoms bonded to one C atom. One can think of this transfer in terms of the rotation of 
the atoms around the double bond and can come up with different reaction paths for the same, such as in-plane and out-of-plane rotations. However, one needs to locate the least energy reaction path on the BOPES that is most likely to be followed by the molecule in real world. 

Keeping in mind the planar structure and symmetry of the molecule, we have chosen the Cl-C=C-Cl torsional angle (internal coordinate) as the reaction coordinate (IRC) that facilitates the interconversion between cis- and trans-geometric isomers. We chose the equilibrium geometry 
calculated by classical HF theory as the initial guess for cis-/trans- configuration. By choosing the C=C double bond as 
the axis of rotation, we applied the rotation operator given by Fig. \ref{fig:Rotation_Operator} to the two C-Cl single bonds to rotate the two Cl atoms in opposite directions by an angle of 15°. In order to maintain the planar structure of the molecule, we rotated the two hydrogen (H) atoms by the same angle in the same direction as the respective Cl atoms such that the H-C-Cl angle remains constant for all iterations. For each torsion angle, one can further relax the H atoms along the C-H bond to optimize the 
geometry and find minimum energy for each iteration.

\begin{figure*}
    \centering
    \includegraphics[width=0.9\textwidth]{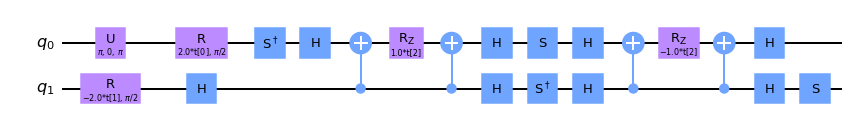}
    \caption{The UCCSD ansatz used to simulate isomerization pathway for \ch{C2H2Cl2}}
    \label{fig:UCCSD_circuit}
\end{figure*}

We used the Hartree-Fock embedding scheme and 
restricted the Active Space to CAS(2,2), choosing only two electrons and two molecular orbitals (corresponding to four spin orbitals) near the Fermi-level for quantum simulation. 
The use of Parity mapping has helped us to reduce the qubit count to two. The ansatz chosen was the UCCSD ansatz with a circuit depth of 15. The circuit diagram has been depicted in Fig. \ref{fig:UCCSD_circuit} 

For the purpose of studying the influence of different basis sets on the isomerization pathways, we have chosen the minimal STO-3G basis and a relatively higher-order 6-31G* basis to compare the energies obtained for each isomer. The results are reported in table \ref{table:VQE_energies_C2H2Cl2} along with classical HF energy data in parenthesis.
\begin{table}[h!]
\centering
\caption{VQE computed energies for cis- and trans- isomers of \ch{C2H2Cl2} in different basis sets}
\label{table:VQE_energies_C2H2Cl2}
\begin{tabular}{|c|c|c|}
\hline
Molecule      & MO Basis & Equilibrium Energy (Hartree) \\ \hline
cis-\ch{C2H2Cl2}   & 6-31G*   & -995.83531 (-995.830114)     \\ \hline
trans-\ch{C2H2Cl2} & 6-31G*   & -995.83402(-995.830492)      \\ \hline
cis-\ch{C2H2Cl2}   & STO-3G   & -985.08824 (-985.07603)      \\ \hline
trans-\ch{C2H2Cl2} & STO-3G   & -985.08892 (-985.07741)      \\ \hline
\end{tabular}
\end{table}

From the table, we observe that there is a very small energy difference between cis-\ch{C2H2Cl2} and trans-\ch{C2H2Cl2}, supporting the fact that dichloroethylene can exist in either of 
its two geometric (stereo) isomeric configurations in nature. This also indicates a possibility of conversion from one isomer to another via rotation about the C=C double bond. Both classical and VQE simulations in STO-3G basis has yielded a 
lower energy for the trans-isomer as compared to the cis-form. However, with 6-31G basis, it has been found that the cis- form has lower energy and is more stable than the trans-form. This suggests that, as discussed in section \ref{molecular hamiltonian}
the use of higher basis sets for the construction of electronic Hamiltonian enhances the accuracy of computation.Moreover, we observe that with the incorporation of single and double excitations of active electrons in our quantum simulations we obtain better energies than the HF energies, which primarily takes into account a single Slater determinant only.

\begin{figure}[H]
\centering
\begin{subfigure}[b]{0.45\textwidth}
   \centering
   \includegraphics[width=1\linewidth]{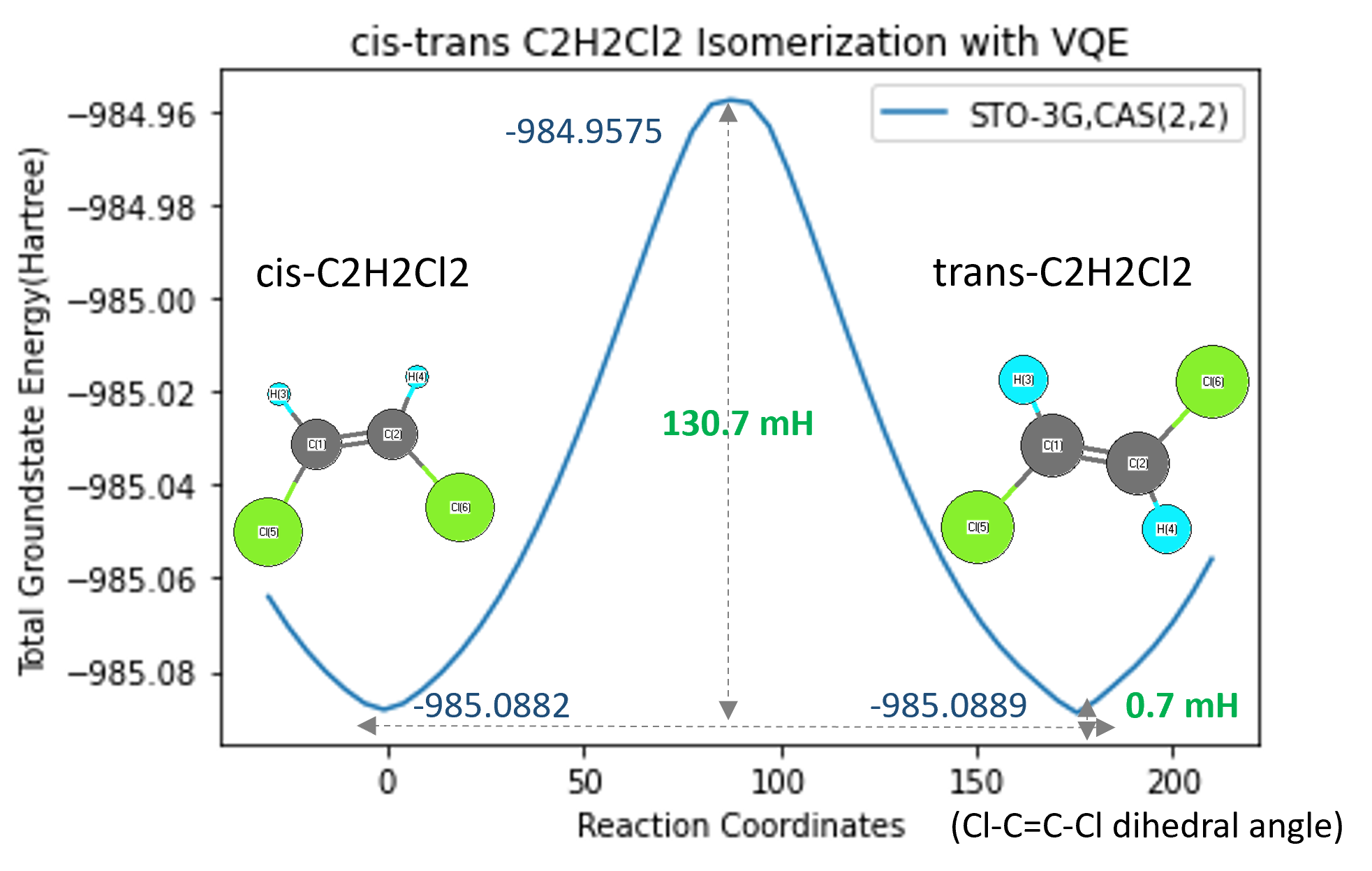}
   \caption{}
   \label{fig:C2H2Cl2 reaction path sto3g} 
\end{subfigure}

\begin{subfigure}[b]{0.45\textwidth}
   \centering
   \includegraphics[width=1\linewidth]{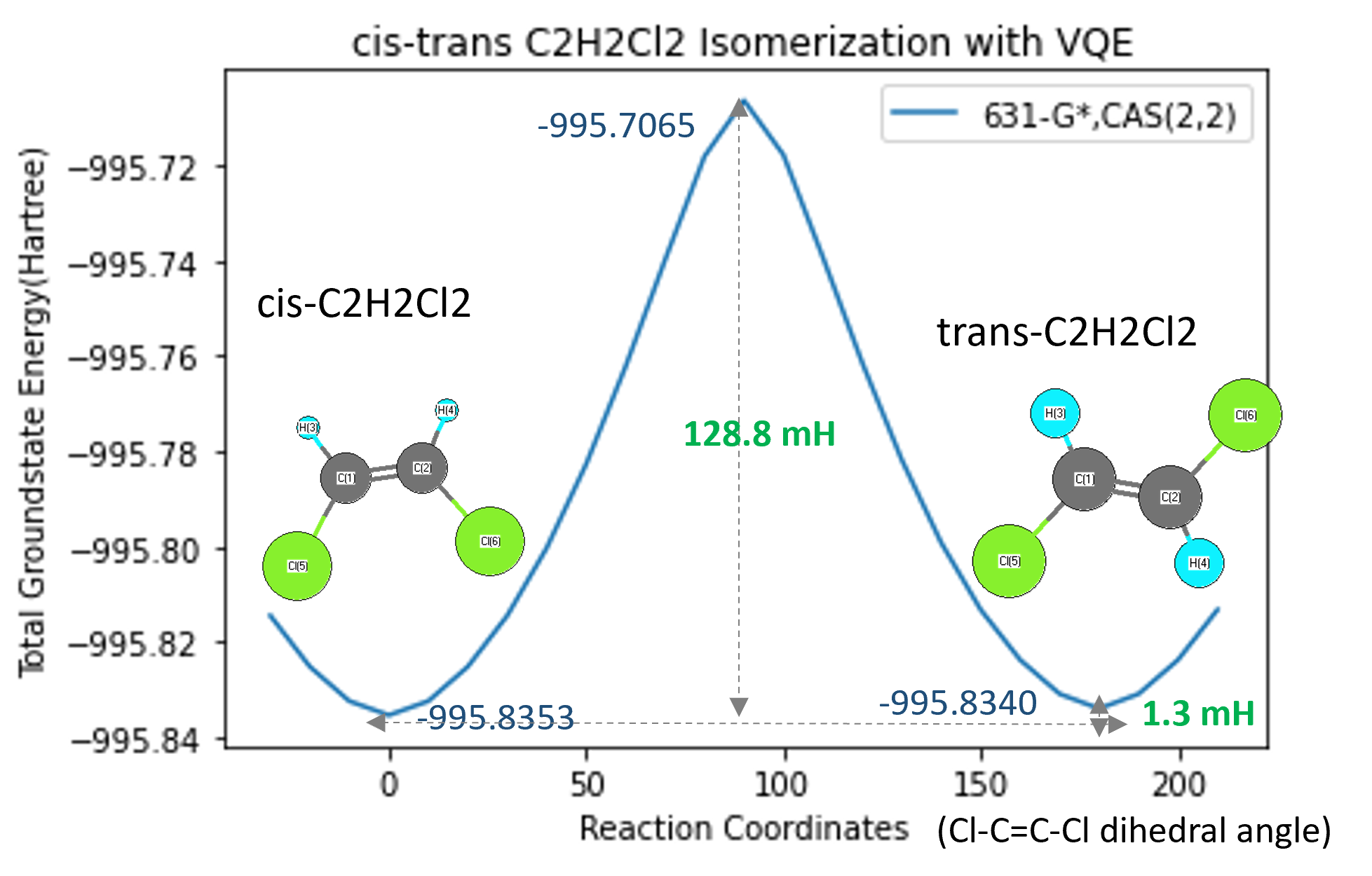}
   \caption{}
   \label{fig:C2H2Cl2 reaction path 631g}
\end{subfigure}

\caption{Isomerization Pathway of C2H2Cl2 obtained with IRC-VQE (a) with STO-3G basis (b) with 6-31G* basis}
\label{fig:C2H2Cl2 isomerization path}
\end{figure}

Fig. \ref{fig:C2H2Cl2 isomerization path} shows the complete reaction paths obtained with the two basis sets. One important observation is that the energy barrier computed between the cis- and trans- forms are on the order of 100 milli-Hartrees. Presence of such large barriers in a simple alkene like dichloroethane eliminates the possibility of interconversion via rotation about double bond and happens through rupture of the double bond.

\subsection{Case Study 2: IRC-VQE applied to \ch{C2H6}} \label{case study 2}
For case study 2, we have considered an example from
Stereochemistry, which deals with the three-dimensional aspects of molecules. Such 3-D aspects are often crucial to understand properties of biological molecules. We have considered a simple alkane - Ethane (\ch{C2H6}), which shows 
conformational isomerism. For alkanes (-C-C-), $\sigma$ bonds are cylindrically symmetrical, allowing “free” rotation around carbon–carbon bonds in open-chain molecules, creating different “conformational isomers” or conformers. 

\begin{figure}[H]
    \centering
    \includegraphics[width=0.45\textwidth]{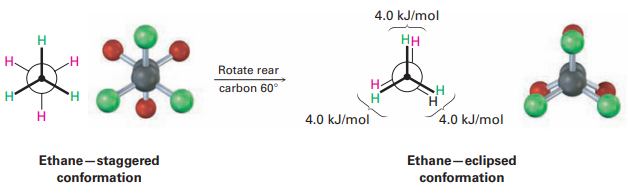}
    \caption{Staggered and Eclipsed Conformers of \ch{C2H6}}
    \label{fig:ethane conformers}
\end{figure}

Experimental studies have predicted that the most stable conformer is the one in which all six hydrogen atoms are located as far as possible from one another. This conformer is referred to as the staggered conformer. The conformer with all 
hydrogen atoms located as close as possible is said to be the least stable one, referred to as the eclipsed conformer and depicted in Fig.\ref{fig:ethane conformers}. Though single C-C bonds facilitates free 
rotation, there is a small (12 kJ/mol) barrier, between staggered and eclipsed conformers of ethane, can be due to torsional strain.

Our objective in this case study was to resolve this tiny rotational energy barrier predicted between the conformers of ethane using the IRC-VQE framework. For this purpose, we have chosen the H-C-C-H torsional angle (internal coordinate) as the reaction coordinate (IRC) that demonstrates the transformation between different conformational isomers. We fixed one set of three hydrogen 
atoms bonded to one carbon atom and rotated the other three hydrogens bonded to the second carbon atom with a step size of 5° varying the H-C-C-H torsional from -180° to +180°. 
Similar to case study 1, we chose the equilibrium geometry calculated by classical HF theory as the initial guess for conformational isomeric configuration. For each torsion 
angle, one can further relax the H atoms along the C-H bond to optimize the geometry and find minimum energy for each iteration.

For probing the influence of different Active space 
transformations on the simulated conformational reaction path, we have chosen CAS (2,2) and CAS (4,4) having 2 electrons and 2 molecular orbitals and 4 electrons and 4 molecular orbitals, respectively. As discussed in case study 1, 
CAS (2,2) resulted in a smaller quantum circuit with 2 qubits and 3 parameters after parity mapping whereas CAS (4,4) yielded a 6-qubit circuit with 26 parameters with the same mapping. This is of particular importance with regard to 
NISQ devices as CAS (2,2) embedding results in quantum circuits which when run on real quantum hardware can possibly produce results that are less corrupted by noise. For more details on the circuits, refer Table \ref{table:Quantum circuit cost with active space size}. For both the 
simulations, we kept the MO basis sets as 6-31G*.

\begin{figure}
\centering
\begin{subfigure}[b]{0.45\textwidth}
   \centering
   \includegraphics[width=1\linewidth]{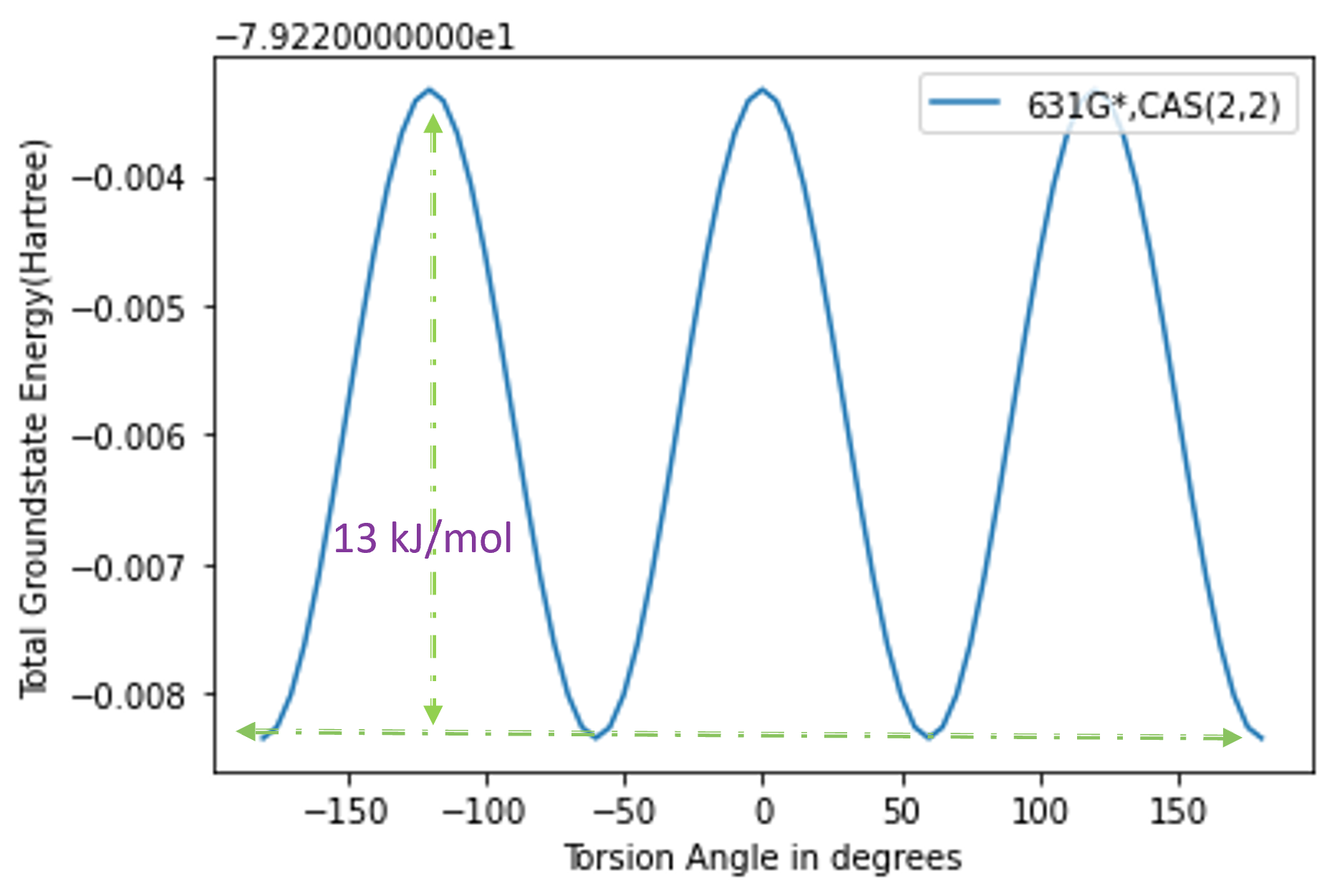}
   \caption{}
   \label{fig:C2H6 reaction path CAS22} 
\end{subfigure}

\begin{subfigure}[b]{0.45\textwidth}
   \centering
   \includegraphics[width=1\linewidth]{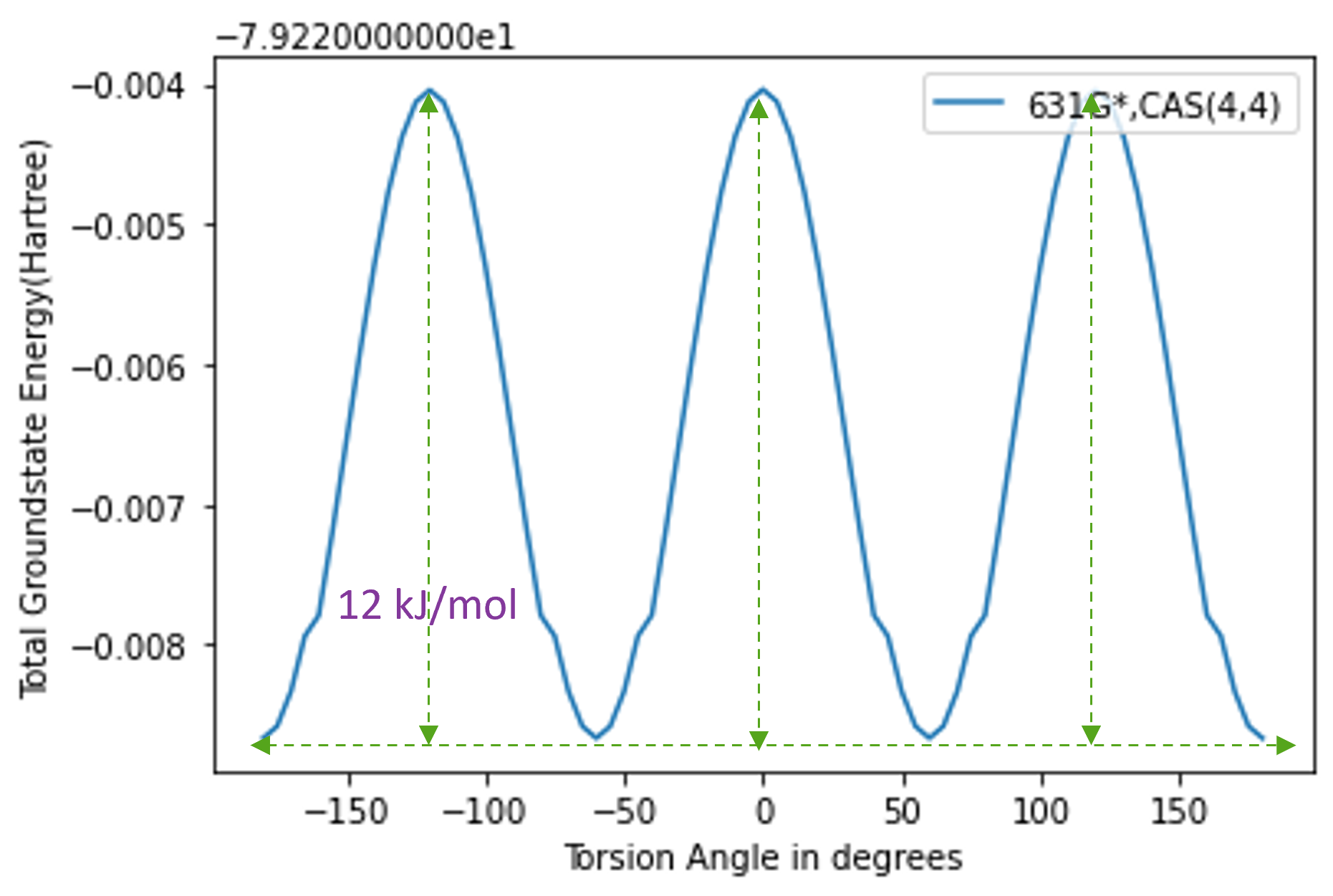}
   \caption{}
   \label{fig:C2H6 reaction path CAS44}
\end{subfigure}

\caption[]{Conformational isomerism observed in \ch{C2H6} using IRC-VQE a) with CAS (2,2) b) CAS (4,4)}
\label{fig:C2H6 isomerization path}
\end{figure}

As evident from the plots in Fig. \ref{fig:C2H6 isomerization path}, the energy barrier obtained with CAS (4,4) has turned out to be close to the experimentally reported value of 12 kJ/mol. This enhancement in accuracy compared to CAS (2,2) came at a higher quantum cost of 6 qubits. This suggests that subjecting more orbitals and electrons for quantum simulation and 
considering their single and double excitations above the reference state can bring about highly accurate computations of energies. It also provides better understanding about the chemical reactions through accurate computation of energy 
barriers and chemical reaction rates.
Define abbreviations and acronyms the first time they are used in the text, even after they have been defined in the abstract. Abbreviations such as 
IEEE, SI, MKS, CGS, ac, dc, and rms do not have to be defined. Do not use abbreviations in the title or heads unless they are unavoidable.

\begin{table}[H]
\centering
\caption{Comparison of quantum cost for different Active Spaces}
\label{table:Quantum circuit cost with active space size}
\begin{tabular}{|c|c|c|c|}
\hline
Active Space & No. Qubits & No. Parameters & Ansatz Depth \\ \hline
CAS (2,2)    & 2          & 3              & 15           \\ \hline
CAS(4,4)     & 6          & 26             & 1647         \\ \hline
\end{tabular}
\end{table}

\begin{table}[H]
\centering
\caption{Energies of conformers of \ch{C2H6} in different Active Spaces and corresponding energy barriers}
\label{table:C2H6 conformer energies}
\begin{tabular}{|c|c|c|c|}
\hline
Conformer                       & Active Space                   & \begin{tabular}[c]{@{}c@{}}Energy\\ (Hartrees)\end{tabular} & \begin{tabular}[c]{@{}c@{}}Barrier\\ (kJ/mol)\end{tabular} \\ \hline
Staggered                       & CAS (2,2)                      & -79.228341                                                  &                                                            \\ \hline
Eclipsed                        & CAS (2,2)                      & -79.223333                                                  & 13.148097                                                  \\ \hline
\multicolumn{1}{|l|}{Staggered} & \multicolumn{1}{l|}{CAS (4,4)} & \multicolumn{1}{l|}{-79.228666}                             & \multicolumn{1}{l|}{}                                      \\ \hline
\multicolumn{1}{|l|}{Eclipsed}  & \multicolumn{1}{l|}{CAS (4,4)} & \multicolumn{1}{l|}{-79.224038}                             & \multicolumn{1}{l|}{12.151007}                             \\ \hline
\end{tabular}
\end{table}

\section{Conclusion and Future Work}
In this work, we present a hybrid quantum-classical 
framework to follow the reaction pathways on the potential energy surfaces of various chemical systems. This framework leverages the enormous potential of near-term and future quantum computers in representing and processing complex 
wavefunctions of large molecular systems and the 
capabilities of conventional classical computers in the most appropriate form to enhance the accuracy and efficiency of quantum chemistry calculations. The generic nature of the framework has been highlighted by describing the various 
choices one could select for each module in the framework depending on the requirements on accuracy and the affordability of the computations. We demonstrate the working of the framework by applying it to trace the cis-trans isomerization reaction path of 1,2-diclorethylene and the conformational isomerization observed in ethane. We 
observe that there is a trade-off between accuracy and quantum resources in the current NISQ era of quantum simulation of matter. 

In future, we plan to improve the scalability of this framework to handle molecules of larger size and larger degrees of freedom. The incorporation of finite-difference based energy gradient could help the classical nonlinear optimizers to trace the reaction paths more accurately than force-field based methods which are approximate. 
Techniques to reduce the quantum cost in terms of lower circuit depth and qubit count needs to be investigated in parallel.

\bibliographystyle{IEEEtran}
\bibliography{refs.bib}

\end{document}